\documentclass[%
reprint,
superscriptaddress,
 amsmath,amssymb,
 aps,
]{revtex4-2}

\usepackage{graphicx}
\usepackage{dcolumn}
\usepackage{bm}
\usepackage{hyperref}
\usepackage{tikz-cd}
\usepackage[compat=1.1.0]{tikz-feynman}
\usepackage{physics}
\usepackage{cancel}
\usepackage{tensor}
\usepackage{simpler-wick}
\usepackage{slashed}

\usepackage{cleveref}
\crefname{equation}{Eq.}{Eqs.}

\newcommand{\degf}{\mathrm{f}}
\newcommand{\str}[1]{\ensuremath{\text{str}\left(#1\right)}}

\allowdisplaybreaks
\hypersetup{breaklinks=true}

\begin{document}

\title{Supergroup Symmetries and the Hierarchy Problem}

\author{Nathaniel Craig}
\affiliation{%
 Department of Physics, University of California, Santa Barbara, California 93106, USA
}%
\affiliation{%
 Kavli Institute for Theoretical Physics, Santa Barbara, CA 93106, USA
}
\author{Emanuele Gendy}%
\affiliation{%
Technische Universit\"{a}t M\"{u}nchen, Physik-Department, 85748 Garching, Germany
}
\author{Jessica N. Howard }%
\affiliation{%
Kavli Institute for Theoretical Physics, Santa Barbara, CA 93106, USA
}
\begin{abstract}

We show that the mass of a scalar field transforming in the fundamental of an $SU(N|N+1)$ supergroup internal symmetry is protected against one-loop corrections from quartic, gauge, and yukawa interactions. Although the negative-norm states responsible for this protection are likely fatal to a unitary interpretation of the theory, such ultraviolet insensitivity may open a new avenue to understanding the lightness of the Higgs.

\end{abstract}

\maketitle

\section{Introduction} \label{sec:intro}

The microscopic origin of the Higgs mass is one of the greatest open questions in particle physics. Within the Standard Model the Higgs mass is an incalculable parameter, while in extensions of the Standard Model it is exquisitely sensitive to physics in the ultraviolet. This raises a puzzle -- the hierarchy problem -- as to why the Higgs is light when the apparent scales of physics beyond the Standard Model (such as the scale of quantum gravity) are far greater. Conventional solutions to the hierarchy problem, such as spacetime supersymmetry and spontaneously broken global symmetries, render the Higgs mass calculable by augmenting the Standard Model with additional symmetry structure. The observed Higgs mass and avoidance of fine tuning in the microscopic theory correspondingly predict new partner particles within an order of magnitude of the electroweak scale. The absence of evidence for these particles has led to both interrogation of the logic behind the hierarchy problem and a proliferation of new approaches to understanding the Higgs mass, ranging from discrete symmetries \cite{Chacko:2005pe} to cosmological selection of the weak scale \cite{Graham:2015cka}. 

In this paper, we explore a novel supersymmetry-based solution to the hierarchy problem -- not in the familiar guise of a spacetime symmetry, but instead as a supergroup internal symmetry. Remarkably, we find at one loop that the mass of a scalar in the fundamental representation of this symmetry does not receive corrections from its own self-coupling, a gauge coupling to a vector multiplet, or a yukawa coupling to a spinor multiplet:
\begin{align}
    \delta m^2_{\Phi
    } = 0 \ .
\end{align}
This type of symmetry has previously been invoked in a range of relativistic contexts, most notably in the $SU(2|1)$ extension of the $SU(2) \times U(1)$ gauge theory of the electroweak interactions \cite{Neeman:1979wp, Fairlie:1979at, Dondi:1979ib, Taylor:1979sm}. Supergroup internal symmetries also play a role in Lagrangian formulations of quenched QCD \cite{Bernard:1992mk}. Spontaneously broken $SU(N|N)$ supergroups have been employed for the Pauli-Villars-like regularization of $SU(N)$ Yang-Mills theory \cite{Arnone:2001iy}, providing inspiration for this work.

There is, of course, an immediate cause for skepticism about a physical interpretation of quantum field theories with supergroup internal symmetries: they generally feature both wrong-sign and wrong-statistics ghost fields. When quantized conventionally, these ghosts lead to catastrophic vacuum instability and causality violation. Indefinite-metric quantization exchanges these fatal flaws for negative-norm states that render the theory non-unitary. Whether a physically sensible theory can be salvaged (perhaps by selection rules or heuristics that prevent positive-norm states from scattering into negative-norm states) is far from clear. These well-known pathologies of non-unitary theories seem to make them irrelevant not just to the hierarchy problem in particular, but to realistic quantum field theories in general. As we will see, however, the surprising radiative properties of theories with supergroup internal symmetries may warrant further exploration. 

Solving naturalness problems with negative-norm states is not without precedent. The Lee-Wick \cite{Lee:1969fy, Lee:1970iw} extension of the Standard Model \cite{Grinstein:2007mp} controls UV corrections to the Higgs potential by invoking wrong-sign higher-derivative interactions that can be interpreted as arising from the exchange of ghost fields. Various prescriptions may rescue unitarity of Lee-Wick models, at least in perturbation theory \cite{Lee:1969fy, Cutkosky:1969fq}. In many respects a supergroup solution to the hierarchy problem most closely resembles Lee-Wick theory, with the considerable advantage of an underlying symmetry that organizes the spectrum of states and ensures genuine UV insensitivity.   Related precedents include the schematic ``Energy $\rightarrow -$ Energy'' symmetry \cite{Kaplan:2005rr} (to control the cosmological constant via an entire Standard Model's worth of ghosts) and agravity \cite{Salvio:2014soa} (to explain the absence of an ultraviolet scale associated with gravity by invoking a ghost partner of the graviton). 

Various attempts have been made to argue for a unitary interpretation of theories with wrong-sign and/or wrong-statistics ghosts (see e.g.~\cite{Lee:1969fy, Lee:1970iw, Cutkosky:1969fq}, and more recently \cite{vanTonder:2006ye, vanTonder:2008ub, Grinstein:2008bg, Anselmi:2017lia, Anselmi:2018kgz, Donoghue:2019fcb}). In this paper our purpose is to explore the radiative structure of theories with supergroup internal symmetries without advocating for particular unitary interpretations. Nonetheless, this radiative structure may motivate further study of such interpretations.

Looking beyond perturbation theory, evidence for the existence of quantum field theories with $SU(N|M)$ internal symmetry is ambiguous. $\mathcal{N} = 4$ supersymmetric $U(N|M)$ supergroup gauge theories arise as the low-energy limit of string theory on $N$ ordinary D-branes and $M$ negative D-branes \cite{Dijkgraaf:2016lym}, although the apparent pathologies associated with negative-tension branes may indicate fatal instabilities. The vacuum geometry and Seiberg-Witten curves have been constructed for $\mathcal{N} = 2$ supersymmetric $SU(N|M)$ supergroup gauge theories, offering suggestive evidence that these theories may be well-defined outside of perturbation theory \cite{Dijkgraaf:2016lym}. More broadly, one might hope that the existence of consistent non-unitary conformal field theories in two dimensions \cite{Cardy:1985yy, Belavin:1984vu, Fisher:1978pf} may guide a physical interpretation of non-unitary field theories in higher dimensions.

This paper is organized as follows: we begin with a brief review of essential properties of $SU(N|M)$ supergroups in Sec.~\ref{sec:supergroups}. In Sec.~\ref{sec:model} we consider a toy model of Lorentz scalars, vectors, and spinors charged under an $SU(N|M)$ supergroup, with a particular eye towards radiative corrections to the mass of Lorentz scalars transforming in the fundamental. We find that all such corrections vanish at one loop for $M = N+1$ when the $SU(N|M)$ symmetry is exact, whereas they are at most logarithmically divergent when the symmetry is softly broken. We briefly discuss aspects of spontaneous symmetry breaking in both global and local $SU(N|M)$ supergroups in Sec.~\ref{sec:ssb}. Conclusions, future directions, and speculations about potential applications to the Standard Model are summarized in Sec.~\ref{sec:conc}. Further details are reserved for a companion paper \cite{toappear}.

\section{Lie Supergroups} \label{sec:supergroups}

To set the stage, we briefly review some of the essential properties of the $\mathfrak{su}(N|M)$ superalgebra and $SU(N|M)$ supergroup (see e.g.~\cite{Bars1984}). In the $N+M$-dimensional fundamental representation the algebra is furnished by matrices of the form
\begin{align}
\mathcal{H}=\begin{pmatrix}
H_N & \theta \\
\theta^\dagger & H_M
\end{pmatrix}\ ,
\end{align}
where $H_{N}$ and $H_{M}$ are hermitian $N\times N$ and $M \times M$ matrices (respectively) with complex bosonic elements satisfying ${\rm tr}(H_N) = {\rm tr} (H_M)$, while $\theta$ is a $N\times M$ matrix of complex, anticommuting Grassmann numbers. The $\mathcal{H}$ are supertraceless in the sense that $\str{\mathcal{H}}\equiv\tr(\sigma_3\mathcal{H})=\tr(H_N)-\tr(H_M) = 0,$ where
	\begin{align}
		\sigma_3=\begin{pmatrix}
			\mathbb{I}_{N\times N} & 0 \\
			0 & -\mathbb{I}_{M\times M}
		\end{pmatrix}\ .
	\end{align}

 In general $\mathcal{H}$ can be assembled from a linear combination of ordinary Lie generators $T^a_N, T^b_M, t_{U}$ parametrizing the $SU(N) \times SU(M) \times U(1)$ bosonic subalgebra (where in a suitable basis $T^a_N$ consists of ordinary $SU(N)$ generators $t^a_N$ in the upper $N \times N$ block and $T^b_M$ consists of ordinary $SU(M)$ generators $t^b_M$ in the lower $M \times M$ block), and supergenerators $S^i, \tilde{S}^i$ that  that close into the ordinary generators under anticommutation,
\begin{align}
S^{i}&=\frac{1}{2}\begin{pmatrix}
			0 & s^{i} \nonumber\\
			s^{\dagger i} &  0
		\end{pmatrix}\ ,&
		\tilde{S}^{i}&=\frac{1}{2}\begin{pmatrix}
			0 & \tilde{s}^{i} \\
			\tilde{s}^{\dagger i} &  0
		\end{pmatrix}\ .\nonumber
	\end{align}
Here $s^i$ are the $NM$, $N\times M$ matrices with $-i$ in one entry and 0 everywhere else, and $\tilde{s}^i$ are the $NM$, $N\times M$ matrices with $1$ in one entry and 0 everywhere else.
Collectively, these generators $\lambda_I \in \{T^a_N, T^b_M, t_{U}, S^i, \tilde{S}^i \}$ can be exponentiated to form elements of the $SU(N|M)$ supergroup. 

The super Killing metric of the group given by $g_{IJ} = 2 \str{\lambda_I \lambda_J}$ takes the schematic form
\begin{align}
g_{IJ} = \left(\begin{array}{c|c}
			\begin{array}{ccc}
				\mathbb{I}_N
				& & \\
				& {\rm sgn}(N-M) &  \\
				& & -\mathbb{I}_M\\
			\end{array}&0\\\hline 
			0&
			\ddots
		\end{array}
		\right)
\end{align}
where the first three diagonal factors respectively correspond to the $SU(N), U(1)$, and $SU(M)$ generators, while the $\dots$ correspond to the supergenerators. The relative signs between the $SU(N)$ and $SU(M)$ factors telegraph the fact that fields in the adjoint will involve wrong-sign ghosts. 

The $\lambda_I$ satisfy a completeness relation
	\begin{align}
		(\lambda_I)_{ij}g^{IJ}(\lambda_J)_{kl}=\frac{1}{2}\left(\delta_{il}\delta_{jk}(-1)^{\degf(j)\degf{(k)}}-\frac{1}{N-M}\delta_{ij}\delta_{kl}\right)\ ,
	\end{align}
	where the grading of an index is given by
	\begin{align}
		\degf{(i)}=\begin{cases}
			0 \qquad\text{       if $1\leq i\leq N$}\\
			1\qquad \text{       if $N+1\leq i\leq M$}
		\end{cases}\ .
	\end{align}
This completeness relation will feature prominently in what follows.
	
\section{An $SU(N|M)$ Toy Model} \label{sec:model}

We now consider Lorentz scalars, spinors, and vectors transforming in various representations of $SU(N|M)$, with a particular eye towards radiative corrections to the masses of Lorentz scalars. In what follows we will almost exclusively be focused on the case $M \neq N$. Although the remarkable radiative properties of pure $SU(N|N)$ Yang-Mills were studied extensively in \cite{Arnone:2001iy}, $SU(N|N)$ does not admit the fundamental representation. 

\subsection{Scalars}

Let us first consider a Lorentz scalar belonging to the fundamental of $SU(N|M)$, which can be written as
	\begin{align}
		\Phi_i=\begin{pmatrix}
			\phi_a\\
			\psi_\alpha
		\end{pmatrix}\ ,
		\label{eq:scalarfieldpar}
	\end{align}
Here $\phi_a$ is a regular $N$-component complex scalar while $\psi_\alpha$ is a $M$-component field of complex Grassmann scalars, i.e. wrong-statistics ghosts. As with all wrong-statistics fields, these ghosts lead to unbounded vacuum energy and causality violation when quantized conventionally, or apparent unitarity violation under indefinite metric quantization.  Supergroup symmetry admits the following renormalizable Lagrangian: 
    \begin{align}
		\mathcal{L}_{\Phi}=\partial_\mu \Phi^{\dagger i}\partial^\mu\Phi_i-m^2\Phi^{\dagger i} \Phi_i - \kappa (\Phi^{\dagger i}\Phi_i)^2\ .
	\end{align}
At one loop, the correction to the mass of the ordinary scalars $\phi_a$ from the quartic interaction takes the form
\begin{align}
\Sigma_\phi = \kappa\left[ 2(N+1)  - 2 M \right] \int\frac{d^dp}{(2\pi)^d}\frac{1}{p^2-m^2}
	\end{align}
where the first term in square brackets comes from the two possible contractions of $\phi_a$ in the loop, while the second term comes from the one possible contraction of $\psi_\alpha$; the relative sign is due to the grading of the $\psi_\alpha$. Clearly this vanishes for $M = N+1$, suggesting remarkable finiteness properties for interacting scalars in the fundamental of $SU(N|N+1)$.

The supergroup symmetry can be softly broken by adding a mass term for the $\psi_\alpha$ ghosts not shared by the ordinary scalars $\phi_a$, 
\begin{align}
            \mathcal{L}_{\Phi}\to\mathcal{L}_{\Phi}-m_{\text{soft}}^2\psi^{\dagger \alpha}\psi_\alpha\ .
\end{align}
The effects of soft breaking are quite analogous to soft breaking of spacetime supersymmetry: at one loop the mass of $\phi_a$ accumulates a correction of the form
        \begin{align} \nonumber 
            \delta m^2_\phi =-2(N+1)\frac{\kappa}{16\pi^2} & \left[m_{\text{soft}}^2\left(1+\log(\frac{\mu^2}{m^2+m_{\text{soft}}^2})\right) \right. \\
            & \left. -m^2\log(1+\frac{m_{\text{soft}}^2}{m^2})\right]
        \end{align}
renormalized in $\overline{MS}$. 
	
\subsection{Vectors}

We next consider vector fields of a local $SU(N|M)$ symmetry; see e.g. \cite{Arnone:2001iy} for extensive discussion. These consist of bosonic vectors $A_\mu^{N,a},$ $A_\mu^{M,b}$, and $A_\mu^\lambda$ transforming in the adjoint of the $SU(N), SU(M)$, and $U(1)$ bosonic subgroups, respectively, and fermionic vectors $B_\mu^i$ transforming in the fundamental of one bosonic subgroup and the complex conjugate of the fundamental under the other. The $B_\mu^i$ are wrong-statistics ghosts on account of their fermionic grading, while the $A_\mu^{M}$ are wrong-sign ghosts owing to the super Killing metric. These fields may be collected into a matrix-valued super-gauge field of the form 
\begin{align}
		\mathcal{A}_\mu=\begin{pmatrix}
			A_{\mu}^{N,a}t^a_N&B_\mu^i(s_1+\tilde{s}_i)\\
			(B_\mu^\dagger)^i(s_1^\dagger+\tilde{s}^\dagger_i)&A_{\mu}^{M,b}t^b_M
		\end{pmatrix}
		+A_\mu^\lambda t_{U}\ ,
		\label{eq:gaugefields}
	\end{align}

The local $SU(N|N)$ supergroup is known to possess remarkable finiteness properties  \cite{Arnone:2001iy}. Here, however, our attention is focused on the radiative corrections from $SU(N|M)$ super-gauge fields on the mass of a complex scalar in the fundamental representation. At one loop, this is proportional to the completeness relation
	\begin{align} \nonumber 
		(\lambda_I)_{ik}g^{IJ}(\lambda_J)_{ki} & =\frac{1}{2}\left(\delta_{ii}\delta_{kk}(-1)^{\degf (k)^2}-\frac{1}{N-M}\delta_{ik}\delta_{ki}\right) \\ 
		&=\frac{1}{2}\delta_{ii}\left((N-M)-\frac{1}{N-M}\right)\ .
	\end{align}
As such, for $M = N+1$ the super-gauge fields do not renormalize the mass of the scalar at one loop. 

In theories with spacetime supersymmetry, soft supersymmetry breaking may split the masses of gauge bosons and gauginos in the same vector multiplet without issue. In contrast, one can show that turning on soft masses for any of the $A_\mu$ or $B_\mu$ fields in a local $SU(N|M)$ supergroup leads to perturbative unitarity violation in at scales $E \sim 4 \pi m_{\rm soft} / g$. While the perils of perturbative unitarity violation are modest compared to the larger unitarity questions posed by the proliferation of ghost fields, they indicate that soft breaking is not perfectly analogous between theories with spacetime supersymmetry and supergroup internal symmetries. Note, however, that components of the super-gauge multiplet may acquire mass from spontaneous symmetry breaking, as we explore in Sec.~\ref{sec:ssb}.

\subsection{Spinors}

Finally, we consider Lorentz spinors in various representations of $SU(N|M)$. For clarity, consider first a theory of two super-spinor fields: $\Theta$ transforming in the adjoint of $SU(N|M)$, and $\xi$ transforming in the fundamental. Both contain ordinary fermions as well as bosonic spinors. These spinors may be coupled to a super-scalar field $\Phi$ in the fundamental representation via Yukawa term of the form
	\begin{align}
		\mathcal{L}_{Yuk}=-y\Phi_i\bar{\xi}_j(\lambda_I)_{ji}\Theta^I+\text{ h.c.}\ ,
	\end{align}
The one-loop correction to the mass for $\Phi$ is again proportional to $(\lambda_I)_{ik}g^{IJ}(\lambda_J)_{ki}$ and vanishes for $M = N+1$. 

We may introduce soft masses for select components of $\Theta$ by splitting the metric $g_{IJ}=g_{IJ}^{(1)}+g_{IJ}^{(2)}$ and adding a term of the form
        \begin{align}
            \mathcal{L}\to\mathcal{L}-m_{\Theta,\text{soft}}\bar{\Theta}^Ig_{IJ}^{(2)}\Theta^J\ .
        \end{align}
For example, giving a soft mass to all of the wrong-statistics components of $\Theta$ means taking $g^{(2)IJ}=g^{(F)IJ}$, where $g_{IJ}^{(F)}$ is a matrix equal to $g_{IJ}$ for the fermionic block and zero elsewhere. The resulting one-loop correction to the mass of $\Phi$ is only logarithmically divergent,
        \begin{align}
            \delta m_{\Phi_i}^2 & =-\frac{y^2}{32 \pi^2}m_{\Theta,\text{soft}}^2\left(3-2\log(\frac{\mu^2}{m_{\Theta,\text{soft}}^2})\right) \\ &
            \times\begin{cases}
            \frac{N+1}{2} \qquad\text{if $\degf(i)=0$}\\
            -\frac{N}{2} \qquad\text{if $\degf(i)=1$}\ .
            \end{cases}
        \end{align}
Note that the cancellation of loop corrections to the $\Phi$ mass does not occur when e.g. $\Theta$ transforms as a singlet and $\xi$ transforms as a fundamental of $SU(N|M)$. However, consider a theory with two spinors $\xi, \tilde{\xi}$ belonging to the fundamental of $SU(N|M)$, plus two spinors $\chi, \tilde{\chi}$ that are singlets of $SU(N|M)$. Both $\xi_i$ and $\chi$ are ordinary super-spinors, while $\tilde{\xi}_i$ and $\tilde{\chi}$ are super-spinors in which the grading has been exchanged between the upper and lower components of the multiplet \cite{Bars1984}. Coupling these fields to the super-scalar field $\Phi$ via
    \begin{align}
        \mathcal{L}_{Yuk}= -y\Phi_i(\bar{\xi}^i\chi+\bar{\tilde{\xi}}^i
        \tilde{\chi})+\text{h.c.}\ .
    \end{align}
  results in a vanishing one-loop correction to the mass of $\Phi$ for super-spinors in the fundamental and singlet representations.

\section{Spontaneous Symmetry Breaking} \label{sec:ssb}

Let us briefly consider the spontaneous breaking of $SU(N|M)$ down to its $SU(N) \times SU(M) \times U(1)$ bosonic subgroup, which may be relevant for phenomenological applications \footnote{Our focus remains on $M \neq N$; for a discussion of spontaneous breaking with $M=N$ see \cite{Arnone:2001iy}.}.Unfortunately, the potential for a Lorentz scalar superfield in the adjoint of $SU(N|M)$ contains unbounded directions at tree level that make it unsuitable for this purpose. One can instead consider a field $\Sigma^i_j$ transforming as a direct product of fundamental and antifundamental representations. For appropriate choices of parameters, the potential for $\Sigma$ features a radiatively stable local minimum in which $\langle \Sigma \rangle = \rho \sigma_3$, so that $SU(N|M) \rightarrow SU(N) \times SU(M) \times U(1)$.  For further details, see \cite{toappear}. Around this minimum, the wrong-statistics components $B_\mu$ of the super-vector field $\mathcal{A}_\mu$ acquire masses $m_B = 2 g \rho$, while both the ordinary vectors $A_\mu^N$ and the wrong-sign ghost vectors $A_\mu^M$ remain massless.

At one loop a super-scalar field $\Phi$ in the fundamental of $SU(N|M)$ acquires a mass proportional to $m_B$, which in the limit $m_\Phi^2 \ll m_B^2$ takes the form
        \begin{align}
			m^2_{\Phi,phys}\approx &m^2_{\Phi}+m_{B}^2\frac{g^2}{16\pi^2}\left(\frac{N+1}{2}\right)\left(1+2\log(\frac{\mu^2}{m^2_B})\right)\ .
		\end{align}
While in principle the wrong-statistics vectors can be decoupled by taking $\rho \rightarrow \infty$, $\Phi$ may not be kept light without fine tuning.

\section{Conclusion} \label{sec:conc}

In this work, we have illustrated the remarkable finiteness properties of theories with global or local $SU(N|N+1)$ supergroup internal symmetries, with a particular focus on vanishing one-loop corrections to the mass of scalars in the fundamental representation from loops of scalars, vectors, and (suitably chosen) spinors. These vanishing corrections owe to the presence of negative-norm states whose existence and couplings are dictated by the symmetries of the theory, in contrast to Lee-Wick models. We have also explored the consequences of softly breaking the supergroup internal symmetry with dimensionful parameters, which are quite analogous to the effects of soft breaking in theories with spacetime supersymmetry. 

Needless to say, there are many possible directions for future work. While vanishing loop corrections from supergroup internal symmetries are amusing, phenomenological applications require a plausible unitary interpretation that is presently lacking. Moreover, we have focused exclusively on one-loop corrections to scalar masses, leaving open the study of additional divergences at one loop and beyond. It is not entirely obvious that the cancellations observed at one loop persist to higher orders, or that soft breaking introduces only logarithmic divergences beyond one loop. Regardless, even the remarkable one-loop properties of these theories motivate renewed exploration of potential unitary interpretations.  

While the unitarity problems surrounding negative-norm states may well prove fatal for supergroups, in the event that unitarity can be salvaged it is tempting to briefly speculate about possible phenomenological applications to the hierarchy problem. In contrast with spacetime supersymmetry, the entire Standard Model need not be embedded into supergroups. Rather, global or local supergroup internal symmetries may exist alongside ordinary internal symmetries, in which case they may mitigate UV sensitivity in selected sectors of the Standard Model and explain why complete solutions to the hierarchy problem do not appear below the TeV scale (thereby solving what is known as the ``little hierarchy problem''). 

Even in the event that the entire Standard Model is embedded into supergroups, such an `internally supersymmetric Standard Model' would not merely recapitulate the features of spacetime supersymmetry (up to exchanging the latter's opposite-spin superpartners with the former's same-spin-but-wrong-statistics counterparts). Internal supergroup symmetries are much less constraining in that (for example) they need not relate the quartic of the Higgs boson to the electroweak gauge couplings as in spacetime supersymmetry, lessening the challenge of naturally explaining the observed Higgs mass.

\section*{Acknowledgements}

We would like to thank Hsin-Chia Cheng, Savas Dimopoulos, Roni Harnik, Jay Hubisz, Graham Kribs, Markus Luty, Surjeet Rajendran, Lisa Randall, John Terning, and Giovanni Villadoro for useful conversations and healthy skepticism. This work was supported in part by the U.S. Department of Energy under the grant DE-SC0011702 and performed in part at the Kavli Institute for Theoretical Physics, supported by the National Science Foundation under Grant No.~NSF PHY-1748958. JNH was supported by the National Science Foundation under Grant No. NSF PHY-1748958 and by the Gordon and Betty Moore Foundation through Grant No. GBMF7392. EG is supported by the Collaborative Research Center SFB1258 and the Excellence Cluster ORIGINS, which is funded by the Deutsche Forschungsgemeinschaft (DFG, German Research Foundation) under Germany’s Excellence Strategy – EXC-2094-390783311.

\bibliography{biblio}

\begin{thebibliography}{27}%
\makeatletter
\providecommand \@ifxundefined [1]{%
 \@ifx{#1\undefined}
}%
\providecommand \@ifnum [1]{%
 \ifnum #1\expandafter \@firstoftwo
 \else \expandafter \@secondoftwo
 \fi
}%
\providecommand \@ifx [1]{%
 \ifx #1\expandafter \@firstoftwo
 \else \expandafter \@secondoftwo
 \fi
}%
\providecommand \natexlab [1]{#1}%
\providecommand \enquote  [1]{``#1''}%
\providecommand \bibnamefont  [1]{#1}%
\providecommand \bibfnamefont [1]{#1}%
\providecommand \citenamefont [1]{#1}%
\providecommand \href@noop [0]{\@secondoftwo}%
\providecommand \href [0]{\begingroup \@sanitize@url \@href}%
\providecommand \@href[1]{\@@startlink{#1}\@@href}%
\providecommand \@@href[1]{\endgroup#1\@@endlink}%
\providecommand \@sanitize@url [0]{\catcode `\\12\catcode `\$12\catcode
  `\&12\catcode `\#12\catcode `\^12\catcode `\_12\catcode `\%12\relax}%
\providecommand \@@startlink[1]{}%
\providecommand \@@endlink[0]{}%
\providecommand \url  [0]{\begingroup\@sanitize@url \@url }%
\providecommand \@url [1]{\endgroup\@href {#1}{\urlprefix }}%
\providecommand \urlprefix  [0]{URL }%
\providecommand \Eprint [0]{\href }%
\providecommand \doibase [0]{https://doi.org/}%
\providecommand \selectlanguage [0]{\@gobble}%
\providecommand \bibinfo  [0]{\@secondoftwo}%
\providecommand \bibfield  [0]{\@secondoftwo}%
\providecommand \translation [1]{[#1]}%
\providecommand \BibitemOpen [0]{}%
\providecommand \bibitemStop [0]{}%
\providecommand \bibitemNoStop [0]{.\EOS\space}%
\providecommand \EOS [0]{\spacefactor3000\relax}%
\providecommand \BibitemShut  [1]{\csname bibitem#1\endcsname}%
\let\auto@bib@innerbib\@empty
\bibitem [{\citenamefont {Chacko}\ \emph {et~al.}(2006)\citenamefont {Chacko},
  \citenamefont {Goh},\ and\ \citenamefont {Harnik}}]{Chacko:2005pe}%
  \BibitemOpen
  \bibfield  {author} {\bibinfo {author} {\bibfnamefont {Z.}~\bibnamefont
  {Chacko}}, \bibinfo {author} {\bibfnamefont {H.-S.}\ \bibnamefont {Goh}},\
  and\ \bibinfo {author} {\bibfnamefont {R.}~\bibnamefont {Harnik}},\
  }\bibfield  {title} {\bibinfo {title} {{The Twin Higgs: Natural electroweak
  breaking from mirror symmetry}},\ }\href
  {https://doi.org/10.1103/PhysRevLett.96.231802} {\bibfield  {journal}
  {\bibinfo  {journal} {Phys. Rev. Lett.}\ }\textbf {\bibinfo {volume} {96}},\
  \bibinfo {pages} {231802} (\bibinfo {year} {2006})},\ \Eprint
  {https://arxiv.org/abs/hep-ph/0506256} {arXiv:hep-ph/0506256} \BibitemShut
  {NoStop}%
\bibitem [{\citenamefont {Graham}\ \emph {et~al.}(2015)\citenamefont {Graham},
  \citenamefont {Kaplan},\ and\ \citenamefont {Rajendran}}]{Graham:2015cka}%
  \BibitemOpen
  \bibfield  {author} {\bibinfo {author} {\bibfnamefont {P.~W.}\ \bibnamefont
  {Graham}}, \bibinfo {author} {\bibfnamefont {D.~E.}\ \bibnamefont {Kaplan}},\
  and\ \bibinfo {author} {\bibfnamefont {S.}~\bibnamefont {Rajendran}},\
  }\bibfield  {title} {\bibinfo {title} {{Cosmological Relaxation of the
  Electroweak Scale}},\ }\href {https://doi.org/10.1103/PhysRevLett.115.221801}
  {\bibfield  {journal} {\bibinfo  {journal} {Phys. Rev. Lett.}\ }\textbf
  {\bibinfo {volume} {115}},\ \bibinfo {pages} {221801} (\bibinfo {year}
  {2015})},\ \Eprint {https://arxiv.org/abs/1504.07551} {arXiv:1504.07551
  [hep-ph]} \BibitemShut {NoStop}%
\bibitem [{\citenamefont {Ne'eman}(1979)}]{Neeman:1979wp}%
  \BibitemOpen
  \bibfield  {author} {\bibinfo {author} {\bibfnamefont {Y.}~\bibnamefont
  {Ne'eman}},\ }\bibfield  {title} {\bibinfo {title} {{Irreducible Gauge Theory
  of a Consolidated Weinberg-Salam Model}},\ }\href
  {https://doi.org/10.1016/0370-2693(79)90521-5} {\bibfield  {journal}
  {\bibinfo  {journal} {Phys. Lett. B}\ }\textbf {\bibinfo {volume} {81}},\
  \bibinfo {pages} {190} (\bibinfo {year} {1979})}\BibitemShut {NoStop}%
\bibitem [{\citenamefont {Fairlie}(1979)}]{Fairlie:1979at}%
  \BibitemOpen
  \bibfield  {author} {\bibinfo {author} {\bibfnamefont {D.~B.}\ \bibnamefont
  {Fairlie}},\ }\bibfield  {title} {\bibinfo {title} {{Higgs' Fields and the
  Determination of the Weinberg Angle}},\ }\href
  {https://doi.org/10.1016/0370-2693(79)90434-9} {\bibfield  {journal}
  {\bibinfo  {journal} {Phys. Lett. B}\ }\textbf {\bibinfo {volume} {82}},\
  \bibinfo {pages} {97} (\bibinfo {year} {1979})}\BibitemShut {NoStop}%
\bibitem [{\citenamefont {Dondi}\ and\ \citenamefont
  {Jarvis}(1979)}]{Dondi:1979ib}%
  \BibitemOpen
  \bibfield  {author} {\bibinfo {author} {\bibfnamefont {P.~H.}\ \bibnamefont
  {Dondi}}\ and\ \bibinfo {author} {\bibfnamefont {P.~D.}\ \bibnamefont
  {Jarvis}},\ }\bibfield  {title} {\bibinfo {title} {{A supersymmetric
  Weinberg-Salam model}},\ }\href
  {https://doi.org/10.1016/0370-2693(79)90652-X} {\bibfield  {journal}
  {\bibinfo  {journal} {Phys. Lett. B}\ }\textbf {\bibinfo {volume} {84}},\
  \bibinfo {pages} {75} (\bibinfo {year} {1979})},\ \bibinfo {note} {[Erratum:
  Phys.Lett.B 87, 403 (1979)]}\BibitemShut {NoStop}%
\bibitem [{\citenamefont {Taylor}(1979)}]{Taylor:1979sm}%
  \BibitemOpen
  \bibfield  {author} {\bibinfo {author} {\bibfnamefont {J.~G.}\ \bibnamefont
  {Taylor}},\ }\bibfield  {title} {\bibinfo {title} {{Electroweak Theory in
  SU(2/1)}},\ }\href {https://doi.org/10.1016/0370-2693(79)91120-1} {\bibfield
  {journal} {\bibinfo  {journal} {Phys. Lett. B}\ }\textbf {\bibinfo {volume}
  {83}},\ \bibinfo {pages} {331} (\bibinfo {year} {1979})}\BibitemShut
  {NoStop}%
\bibitem [{\citenamefont {Bernard}\ and\ \citenamefont
  {Golterman}(1992)}]{Bernard:1992mk}%
  \BibitemOpen
  \bibfield  {author} {\bibinfo {author} {\bibfnamefont {C.~W.}\ \bibnamefont
  {Bernard}}\ and\ \bibinfo {author} {\bibfnamefont {M.~F.~L.}\ \bibnamefont
  {Golterman}},\ }\bibfield  {title} {\bibinfo {title} {{Chiral perturbation
  theory for the quenched approximation of QCD}},\ }\href
  {https://doi.org/10.1103/PhysRevD.46.853} {\bibfield  {journal} {\bibinfo
  {journal} {Phys. Rev. D}\ }\textbf {\bibinfo {volume} {46}},\ \bibinfo
  {pages} {853} (\bibinfo {year} {1992})},\ \Eprint
  {https://arxiv.org/abs/hep-lat/9204007} {arXiv:hep-lat/9204007} \BibitemShut
  {NoStop}%
\bibitem [{\citenamefont {Arnone}\ \emph {et~al.}(2002)\citenamefont {Arnone},
  \citenamefont {Kubyshin}, \citenamefont {Morris},\ and\ \citenamefont
  {Tighe}}]{Arnone:2001iy}%
  \BibitemOpen
  \bibfield  {author} {\bibinfo {author} {\bibfnamefont {S.}~\bibnamefont
  {Arnone}}, \bibinfo {author} {\bibfnamefont {Y.~A.}\ \bibnamefont
  {Kubyshin}}, \bibinfo {author} {\bibfnamefont {T.~R.}\ \bibnamefont
  {Morris}},\ and\ \bibinfo {author} {\bibfnamefont {J.~F.}\ \bibnamefont
  {Tighe}},\ }\bibfield  {title} {\bibinfo {title} {{Gauge invariant
  regularization via SU(N|N)}},\ }\href
  {https://doi.org/10.1142/S0217751X02009722} {\bibfield  {journal} {\bibinfo
  {journal} {Int. J. Mod. Phys. A}\ }\textbf {\bibinfo {volume} {17}},\
  \bibinfo {pages} {2283} (\bibinfo {year} {2002})},\ \Eprint
  {https://arxiv.org/abs/hep-th/0106258} {arXiv:hep-th/0106258} \BibitemShut
  {NoStop}%
\bibitem [{\citenamefont {Lee}\ and\ \citenamefont {Wick}(1969)}]{Lee:1969fy}%
  \BibitemOpen
  \bibfield  {author} {\bibinfo {author} {\bibfnamefont {T.~D.}\ \bibnamefont
  {Lee}}\ and\ \bibinfo {author} {\bibfnamefont {G.~C.}\ \bibnamefont {Wick}},\
  }\bibfield  {title} {\bibinfo {title} {{Negative Metric and the Unitarity of
  the S Matrix}},\ }\href {https://doi.org/10.1016/0550-3213(69)90098-4}
  {\bibfield  {journal} {\bibinfo  {journal} {Nucl. Phys. B}\ }\textbf
  {\bibinfo {volume} {9}},\ \bibinfo {pages} {209} (\bibinfo {year}
  {1969})}\BibitemShut {NoStop}%
\bibitem [{\citenamefont {Lee}\ and\ \citenamefont {Wick}(1970)}]{Lee:1970iw}%
  \BibitemOpen
  \bibfield  {author} {\bibinfo {author} {\bibfnamefont {T.~D.}\ \bibnamefont
  {Lee}}\ and\ \bibinfo {author} {\bibfnamefont {G.~C.}\ \bibnamefont {Wick}},\
  }\bibfield  {title} {\bibinfo {title} {{Finite Theory of Quantum
  Electrodynamics}},\ }\href {https://doi.org/10.1103/PhysRevD.2.1033}
  {\bibfield  {journal} {\bibinfo  {journal} {Phys. Rev. D}\ }\textbf {\bibinfo
  {volume} {2}},\ \bibinfo {pages} {1033} (\bibinfo {year} {1970})}\BibitemShut
  {NoStop}%
\bibitem [{\citenamefont {Grinstein}\ \emph {et~al.}(2008)\citenamefont
  {Grinstein}, \citenamefont {O'Connell},\ and\ \citenamefont
  {Wise}}]{Grinstein:2007mp}%
  \BibitemOpen
  \bibfield  {author} {\bibinfo {author} {\bibfnamefont {B.}~\bibnamefont
  {Grinstein}}, \bibinfo {author} {\bibfnamefont {D.}~\bibnamefont
  {O'Connell}},\ and\ \bibinfo {author} {\bibfnamefont {M.~B.}\ \bibnamefont
  {Wise}},\ }\bibfield  {title} {\bibinfo {title} {{The Lee-Wick standard
  model}},\ }\href {https://doi.org/10.1103/PhysRevD.77.025012} {\bibfield
  {journal} {\bibinfo  {journal} {Phys. Rev. D}\ }\textbf {\bibinfo {volume}
  {77}},\ \bibinfo {pages} {025012} (\bibinfo {year} {2008})},\ \Eprint
  {https://arxiv.org/abs/0704.1845} {arXiv:0704.1845 [hep-ph]} \BibitemShut
  {NoStop}%
\bibitem [{\citenamefont {Cutkosky}\ \emph {et~al.}(1969)\citenamefont
  {Cutkosky}, \citenamefont {Landshoff}, \citenamefont {Olive},\ and\
  \citenamefont {Polkinghorne}}]{Cutkosky:1969fq}%
  \BibitemOpen
  \bibfield  {author} {\bibinfo {author} {\bibfnamefont {R.~E.}\ \bibnamefont
  {Cutkosky}}, \bibinfo {author} {\bibfnamefont {P.~V.}\ \bibnamefont
  {Landshoff}}, \bibinfo {author} {\bibfnamefont {D.~I.}\ \bibnamefont
  {Olive}},\ and\ \bibinfo {author} {\bibfnamefont {J.~C.}\ \bibnamefont
  {Polkinghorne}},\ }\bibfield  {title} {\bibinfo {title} {{A non-analytic S
  matrix}},\ }\href {https://doi.org/10.1016/0550-3213(69)90169-2} {\bibfield
  {journal} {\bibinfo  {journal} {Nucl. Phys. B}\ }\textbf {\bibinfo {volume}
  {12}},\ \bibinfo {pages} {281} (\bibinfo {year} {1969})}\BibitemShut
  {NoStop}%
\bibitem [{\citenamefont {Kaplan}\ and\ \citenamefont
  {Sundrum}(2006)}]{Kaplan:2005rr}%
  \BibitemOpen
  \bibfield  {author} {\bibinfo {author} {\bibfnamefont {D.~E.}\ \bibnamefont
  {Kaplan}}\ and\ \bibinfo {author} {\bibfnamefont {R.}~\bibnamefont
  {Sundrum}},\ }\bibfield  {title} {\bibinfo {title} {{A Symmetry for the
  cosmological constant}},\ }\href
  {https://doi.org/10.1088/1126-6708/2006/07/042} {\bibfield  {journal}
  {\bibinfo  {journal} {JHEP}\ }\textbf {\bibinfo {volume} {07}},\ \bibinfo
  {pages} {042}},\ \Eprint {https://arxiv.org/abs/hep-th/0505265}
  {arXiv:hep-th/0505265} \BibitemShut {NoStop}%
\bibitem [{\citenamefont {Salvio}\ and\ \citenamefont
  {Strumia}(2014)}]{Salvio:2014soa}%
  \BibitemOpen
  \bibfield  {author} {\bibinfo {author} {\bibfnamefont {A.}~\bibnamefont
  {Salvio}}\ and\ \bibinfo {author} {\bibfnamefont {A.}~\bibnamefont
  {Strumia}},\ }\bibfield  {title} {\bibinfo {title} {{Agravity}},\ }\href
  {https://doi.org/10.1007/JHEP06(2014)080} {\bibfield  {journal} {\bibinfo
  {journal} {JHEP}\ }\textbf {\bibinfo {volume} {06}},\ \bibinfo {pages}
  {080}},\ \Eprint {https://arxiv.org/abs/1403.4226} {arXiv:1403.4226 [hep-ph]}
  \BibitemShut {NoStop}%
\bibitem [{\citenamefont {van Tonder}(2007)}]{vanTonder:2006ye}%
  \BibitemOpen
  \bibfield  {author} {\bibinfo {author} {\bibfnamefont {A.}~\bibnamefont {van
  Tonder}},\ }\bibfield  {title} {\bibinfo {title} {{Non-perturbative
  quantization of phantom and ghost theories: Relating definite and indefinite
  representations}},\ }\href {https://doi.org/10.1142/S0217751X07036580}
  {\bibfield  {journal} {\bibinfo  {journal} {Int. J. Mod. Phys. A}\ }\textbf
  {\bibinfo {volume} {22}},\ \bibinfo {pages} {2563} (\bibinfo {year}
  {2007})},\ \Eprint {https://arxiv.org/abs/hep-th/0610185}
  {arXiv:hep-th/0610185} \BibitemShut {NoStop}%
\bibitem [{\citenamefont {van Tonder}(2008)}]{vanTonder:2008ub}%
  \BibitemOpen
  \bibfield  {author} {\bibinfo {author} {\bibfnamefont {A.}~\bibnamefont {van
  Tonder}},\ }\bibfield  {title} {\bibinfo {title} {{Unitarity, Lorentz
  invariance and causality in Lee-Wick theories: An Asymptotically safe
  completion of QED}},\ }\href@noop {} {\  (\bibinfo {year} {2008})},\ \Eprint
  {https://arxiv.org/abs/0810.1928} {arXiv:0810.1928 [hep-th]} \BibitemShut
  {NoStop}%
\bibitem [{\citenamefont {Grinstein}\ \emph {et~al.}(2009)\citenamefont
  {Grinstein}, \citenamefont {O'Connell},\ and\ \citenamefont
  {Wise}}]{Grinstein:2008bg}%
  \BibitemOpen
  \bibfield  {author} {\bibinfo {author} {\bibfnamefont {B.}~\bibnamefont
  {Grinstein}}, \bibinfo {author} {\bibfnamefont {D.}~\bibnamefont
  {O'Connell}},\ and\ \bibinfo {author} {\bibfnamefont {M.~B.}\ \bibnamefont
  {Wise}},\ }\bibfield  {title} {\bibinfo {title} {{Causality as an emergent
  macroscopic phenomenon: The Lee-Wick O(N) model}},\ }\href
  {https://doi.org/10.1103/PhysRevD.79.105019} {\bibfield  {journal} {\bibinfo
  {journal} {Phys. Rev. D}\ }\textbf {\bibinfo {volume} {79}},\ \bibinfo
  {pages} {105019} (\bibinfo {year} {2009})},\ \Eprint
  {https://arxiv.org/abs/0805.2156} {arXiv:0805.2156 [hep-th]} \BibitemShut
  {NoStop}%
\bibitem [{\citenamefont {Anselmi}\ and\ \citenamefont
  {Piva}(2017)}]{Anselmi:2017lia}%
  \BibitemOpen
  \bibfield  {author} {\bibinfo {author} {\bibfnamefont {D.}~\bibnamefont
  {Anselmi}}\ and\ \bibinfo {author} {\bibfnamefont {M.}~\bibnamefont {Piva}},\
  }\bibfield  {title} {\bibinfo {title} {{Perturbative unitarity of Lee-Wick
  quantum field theory}},\ }\href {https://doi.org/10.1103/PhysRevD.96.045009}
  {\bibfield  {journal} {\bibinfo  {journal} {Phys. Rev. D}\ }\textbf {\bibinfo
  {volume} {96}},\ \bibinfo {pages} {045009} (\bibinfo {year} {2017})},\
  \Eprint {https://arxiv.org/abs/1703.05563} {arXiv:1703.05563 [hep-th]}
  \BibitemShut {NoStop}%
\bibitem [{\citenamefont {Anselmi}(2018)}]{Anselmi:2018kgz}%
  \BibitemOpen
  \bibfield  {author} {\bibinfo {author} {\bibfnamefont {D.}~\bibnamefont
  {Anselmi}},\ }\bibfield  {title} {\bibinfo {title} {{Fakeons And Lee-Wick
  Models}},\ }\href {https://doi.org/10.1007/JHEP02(2018)141} {\bibfield
  {journal} {\bibinfo  {journal} {JHEP}\ }\textbf {\bibinfo {volume} {02}},\
  \bibinfo {pages} {141}},\ \Eprint {https://arxiv.org/abs/1801.00915}
  {arXiv:1801.00915 [hep-th]} \BibitemShut {NoStop}%
\bibitem [{\citenamefont {Donoghue}\ and\ \citenamefont
  {Menezes}(2019)}]{Donoghue:2019fcb}%
  \BibitemOpen
  \bibfield  {author} {\bibinfo {author} {\bibfnamefont {J.~F.}\ \bibnamefont
  {Donoghue}}\ and\ \bibinfo {author} {\bibfnamefont {G.}~\bibnamefont
  {Menezes}},\ }\bibfield  {title} {\bibinfo {title} {{Unitarity, stability and
  loops of unstable ghosts}},\ }\href
  {https://doi.org/10.1103/PhysRevD.100.105006} {\bibfield  {journal} {\bibinfo
   {journal} {Phys. Rev. D}\ }\textbf {\bibinfo {volume} {100}},\ \bibinfo
  {pages} {105006} (\bibinfo {year} {2019})},\ \Eprint
  {https://arxiv.org/abs/1908.02416} {arXiv:1908.02416 [hep-th]} \BibitemShut
  {NoStop}%
\bibitem [{\citenamefont {Dijkgraaf}\ \emph {et~al.}(2018)\citenamefont
  {Dijkgraaf}, \citenamefont {Heidenreich}, \citenamefont {Jefferson},\ and\
  \citenamefont {Vafa}}]{Dijkgraaf:2016lym}%
  \BibitemOpen
  \bibfield  {author} {\bibinfo {author} {\bibfnamefont {R.}~\bibnamefont
  {Dijkgraaf}}, \bibinfo {author} {\bibfnamefont {B.}~\bibnamefont
  {Heidenreich}}, \bibinfo {author} {\bibfnamefont {P.}~\bibnamefont
  {Jefferson}},\ and\ \bibinfo {author} {\bibfnamefont {C.}~\bibnamefont
  {Vafa}},\ }\bibfield  {title} {\bibinfo {title} {{Negative Branes,
  Supergroups and the Signature of Spacetime}},\ }\href
  {https://doi.org/10.1007/JHEP02(2018)050} {\bibfield  {journal} {\bibinfo
  {journal} {JHEP}\ }\textbf {\bibinfo {volume} {02}},\ \bibinfo {pages}
  {050}},\ \Eprint {https://arxiv.org/abs/1603.05665} {arXiv:1603.05665
  [hep-th]} \BibitemShut {NoStop}%
\bibitem [{\citenamefont {Cardy}(1985)}]{Cardy:1985yy}%
  \BibitemOpen
  \bibfield  {author} {\bibinfo {author} {\bibfnamefont {J.~L.}\ \bibnamefont
  {Cardy}},\ }\bibfield  {title} {\bibinfo {title} {{Conformal Invariance and
  the Yang-lee Edge Singularity in Two-dimensions}},\ }\href
  {https://doi.org/10.1103/PhysRevLett.54.1354} {\bibfield  {journal} {\bibinfo
   {journal} {Phys. Rev. Lett.}\ }\textbf {\bibinfo {volume} {54}},\ \bibinfo
  {pages} {1354} (\bibinfo {year} {1985})}\BibitemShut {NoStop}%
\bibitem [{\citenamefont {Belavin}\ \emph {et~al.}(1984)\citenamefont
  {Belavin}, \citenamefont {Polyakov},\ and\ \citenamefont
  {Zamolodchikov}}]{Belavin:1984vu}%
  \BibitemOpen
  \bibfield  {author} {\bibinfo {author} {\bibfnamefont {A.~A.}\ \bibnamefont
  {Belavin}}, \bibinfo {author} {\bibfnamefont {A.~M.}\ \bibnamefont
  {Polyakov}},\ and\ \bibinfo {author} {\bibfnamefont {A.~B.}\ \bibnamefont
  {Zamolodchikov}},\ }\bibfield  {title} {\bibinfo {title} {{Infinite Conformal
  Symmetry in Two-Dimensional Quantum Field Theory}},\ }\href
  {https://doi.org/10.1016/0550-3213(84)90052-X} {\bibfield  {journal}
  {\bibinfo  {journal} {Nucl. Phys. B}\ }\textbf {\bibinfo {volume} {241}},\
  \bibinfo {pages} {333} (\bibinfo {year} {1984})}\BibitemShut {NoStop}%
\bibitem [{\citenamefont {Fisher}(1978)}]{Fisher:1978pf}%
  \BibitemOpen
  \bibfield  {author} {\bibinfo {author} {\bibfnamefont {M.~E.}\ \bibnamefont
  {Fisher}},\ }\bibfield  {title} {\bibinfo {title} {{Yang-Lee Edge Singularity
  and phi**3 Field Theory}},\ }\href
  {https://doi.org/10.1103/PhysRevLett.40.1610} {\bibfield  {journal} {\bibinfo
   {journal} {Phys. Rev. Lett.}\ }\textbf {\bibinfo {volume} {40}},\ \bibinfo
  {pages} {1610} (\bibinfo {year} {1978})}\BibitemShut {NoStop}%
\bibitem [{\citenamefont {Craig}\ \emph {et~al.}(2024)\citenamefont {Craig},
  \citenamefont {Gendy},\ and\ \citenamefont {Howard}}]{toappear}%
  \BibitemOpen
  \bibfield  {author} {\bibinfo {author} {\bibfnamefont {N.}~\bibnamefont
  {Craig}}, \bibinfo {author} {\bibfnamefont {E.}~\bibnamefont {Gendy}},\ and\
  \bibinfo {author} {\bibfnamefont {J.~N.}\ \bibnamefont {Howard}},\ }\bibfield
   {title} {\bibinfo {title} {to appear},\ }\href@noop {} {\  (\bibinfo {year}
  {2024})}\BibitemShut {NoStop}%
\bibitem [{\citenamefont {Bars}(1984)}]{Bars1984}%
  \BibitemOpen
  \bibfield  {author} {\bibinfo {author} {\bibfnamefont {I.}~\bibnamefont
  {Bars}},\ }\bibinfo {title} {Supergroups and their representations},\ in\
  \href@noop {} {\emph {\bibinfo {booktitle} {Introduction to Supersymmetry in
  Particle and Nuclear Physics}}},\ \bibinfo {editor} {edited by\ \bibinfo
  {editor} {\bibfnamefont {O.}~\bibnamefont {Casta{\~{n}}os}}, \bibinfo
  {editor} {\bibfnamefont {A.}~\bibnamefont {Frank}},\ and\ \bibinfo {editor}
  {\bibfnamefont {L.}~\bibnamefont {Urrutia}}}\ (\bibinfo  {publisher}
  {Springer US},\ \bibinfo {address} {Boston, MA},\ \bibinfo {year} {1984})\
  pp.\ \bibinfo {pages} {107--184}\BibitemShut {NoStop}%
\bibitem [{Note1()}]{Note1}%
  \BibitemOpen
  \bibinfo {note} {Our focus remains on $M \protect \neq N$; for a discussion
  of spontaneous breaking with $M=N$ see \cite {Arnone:2001iy}.}\BibitemShut
  {Stop}%
\end{thebibliography}%

\clearpage
\widetext
\setcounter{figure}{0}
\renewcommand\thefigure{S\arabic{figure}} 

\setcounter{equation}{0}
\renewcommand\theequation{S\arabic{equation}} 
 
 \end{document}